\documentclass[12pt]{article}
\usepackage{amssymb}

\usepackage{graphicx}

\ifx\pdfoutput\undefined
    \relax
\else
    \usepackage{epstopdf}
\fi

\makeatletter
\def\numberbysection{\@addtoreset{equation}{section}
 	\def\theequation{\thesection.\arabic{equation}}}
\makeatother

\numberbysection

\newcommand{\be}{\begin{eqnarray}}
\newcommand{\ee}{\end{eqnarray}}
\newcommand{\non}{\nonumber}
\newcommand{\tr}{\mathop{\rm tr}\nolimits}
\newcommand{\id}{\mathbb{I}}
\newcommand{\ch}{\mathop{\rm cosh}\nolimits}
\newcommand{\sh}{\mathop{\rm sinh}\nolimits}
\newcommand{\tnh}{\mathop{\rm tanh}\nolimits}
\newcommand{\cth}{\mathop{\rm coth}\nolimits}
\newcommand{\csch}{\mathop{\rm cosech}\nolimits}
\newcommand{\sech}{\mathop{\rm sech}\nolimits}

\newcommand{\M}{\mathop{\cal M}\nolimits}

\newcommand{\hh}{\mathop{\mathcal H}\nolimits}

\begin{document}

\begin{titlepage}
\strut\hfill UMTG--250
\vspace{.5in}
\begin{center}

\LARGE Exact solution of the open XXZ chain\\
\LARGE with general integrable boundary terms\\
\LARGE at roots of unity\\[1.0in]
\large Rajan Murgan, Rafael I. Nepomechie and Chi Shi\\[0.8in]
\large Physics Department, P.O. Box 248046, University of Miami\\[0.2in]  
\large Coral Gables, FL 33124 USA\\

\end{center}

\vspace{.5in}

\begin{abstract}
    We propose a Bethe-Ansatz-type solution of the open spin-$1/2$
    integrable XXZ quantum spin chain with {\it general} integrable
    boundary terms and bulk anisotropy values $i \pi/(p+1)$, where $p$
    is a positive integer.  All six boundary parameters are arbitrary,
    and need not satisfy any constraint.  The solution is in terms of
    generalized $T - Q$ equations, having more than one $Q$ function.
    We find numerical evidence that this solution gives the complete
    set of $2^{N}$ transfer matrix eigenvalues, where $N$ is the
    number of spins.
\end{abstract}
\end{titlepage}

\setcounter{footnote}{0}

\section{Introduction}\label{sec:intro}

Although the closed (periodic boundary conditions) spin-${1\over
2}$ XXZ quantum spin chain was solved many years ago \cite{Be, Or, ABA}, the 
corresponding open chain with general integrable boundary terms has 
remained unsolved. Nevertheless, much progress has been achieved on 
this fundamental problem. The special case of diagonal boundary terms 
was solved by Gaudin \cite{Ga} and by Alcaraz {\it et al.} \cite{ABBBQ}.
Sklyanin \cite{Sk} constructed the commuting transfer matrix in terms 
of solutions of the bulk and boundary \cite{Ch} Yang-Baxter equations.
The general solution by de Vega and Gonz\'alez-Ruiz \cite{dVGR} and by
Ghoshal and Zamolodchikov \cite{GZ} of the boundary Yang-Baxter
equation led directly, through Sklyanin's construction, to the general
integrable Hamiltonian \footnote{As discussed further in Section \ref{sec:transfer}, 
here we use a parametrization of the boundary interactions which differs 
from that used in \cite{dVGR, GZ}.}
\be
\hh &=& \hh_{0}
+ {1\over 2}\sh \eta \Big[ 
\cth \alpha_{-} \tnh \beta_{-}\sigma_{1}^{z}
+ \csch \alpha_{-} \sech \beta_{-}\big( 
\ch \theta_{-}\sigma_{1}^{x} 
+ i\sh \theta_{-}\sigma_{1}^{y} \big) \non \\
& & \quad -\cth \alpha_{+} \tnh \beta_{+} \sigma_{N}^{z}
+ \csch \alpha_{+} \sech \beta_{+}\big( 
\ch \theta_{+}\sigma_{N}^{x}
+ i\sh \theta_{+}\sigma_{N}^{y} \big)
\Big]  \,, \label{Hamiltonian} 
\ee
where the ``bulk'' Hamiltonian is given by
\be 
\hh_{0} = {1\over 2}\sum_{n=1}^{N-1}\left( 
\sigma_{n}^{x}\sigma_{n+1}^{x}+\sigma_{n}^{y}\sigma_{n+1}^{y}
+\ch \eta\ \sigma_{n}^{z}\sigma_{n+1}^{z}\right) \,. 
\label{bulkHamiltonian}
\ee
Here $\sigma^{x} \,, \sigma^{y} \,, \sigma^{z}$ are the
standard Pauli matrices, $\eta$ is the bulk anisotropy parameter,
$\alpha_{\pm} \,, \beta_{\pm} \,, \theta_{\pm}$ are arbitrary boundary
parameters, and $N$ is the number of spins. Note that the boundary 
interactions include nondiagonal terms (proportional to $\sigma^{x}$ 
and $\sigma^{y}$), which originate from corresponding nondiagonal terms
in the solutions \cite{dVGR, GZ} of the boundary Yang-Baxter equation.
Due to the presence of these nondiagonal boundary terms, the 
Hamiltonian does not have a simple reference (pseudovacuum) state. 
Hence, many of the standard techniques for solving integrable models 
cannot be applied to this model.

A first step toward solving the case of nondiagonal boundary terms 
was taken in \cite{Ne1}. There it was found that, for bulk anisotropy 
values
\be
\eta = {i \pi\over p+1}\,, \qquad p= 1 \,, 2 \,, \ldots \,,
\label{etavalues}
\ee
(and hence $q \equiv e^{\eta}$ is a root of unity, satisfying
$q^{p+1}=-1$) and arbitrary values of the boundary parameters, the
transfer matrix $t(u)$ (see Section \ref{sec:transfer}) obeys a
functional relation of order $p+1$.  For example, the first three
functional relations are given by
\be
&p=1:& \qquad  t(u)\ t(u +\eta) - \delta(u) - \delta(u+\eta)  = 
f(u) \,, \label{p1functrltn}  \\
&p=2:& \qquad t(u)\ t(u+\eta)\ t(u+2\eta)  -  \delta(u)\ t(u +2\eta)
- \delta(u+\eta)\ t(u)  \non  \\
& & \qquad \qquad  -  \delta(u+2\eta)\ t(u +\eta)  = f(u) \,,  \label{p2functrltn}  \\
&p=3:& \qquad t(u)\ t(u +\eta)\ t(u +2\eta)\ t(u +3\eta) 
- \delta(u+3\eta)\ t(u +\eta)\ t(u +2\eta) \non  \\
& & \qquad \qquad - \delta(u)\ t(u +2\eta)\ t(u +3\eta) 
- \delta(u+\eta)\ t(u)\ t(u +3\eta)    \label{p3functrltn}  \\
& & \qquad \qquad 
- \delta(u+2\eta)\ t(u)\ t(u +\eta) 
+ \delta(u+\eta)\ \delta(u+3\eta) + \delta(u)\ \delta(u+2\eta)
= f(u) \,, \non 
\ee 
where $\delta(u)$ and $f(u)$ are scalar functions which depend on the
boundary parameters, whose explicit expressions are given in Appendix
\ref{sec:expressions}.  (Similar results had been known for closed
RSOS models \cite{BP, BR} and closed spin chains \cite{Ba1, BLZ,
KSS}.)  Expressions for the eigenvalues of the transfer matrix were
also proposed in \cite{Ne1}.  However, since these expressions are
rather complicated and are in terms of zeros of the eigenvalues,
rather than zeros of a $Q$ function (as in conventional Bethe Ansatz),
they are probably not very useful.

By exploiting these functional relations \cite{Ne2} as well as by
other means \cite{CLSW, NR, YNZ}, a more conventional Bethe Ansatz
solution was then found for arbitrary values of the bulk anisotropy,
provided that the boundary parameters obey the constraint
\be
\alpha_- + \epsilon_1 \beta_- + \epsilon_2 \alpha_+ + 
\epsilon_3 \beta_+ = \epsilon_0 (\theta_- -\theta_+) 
+\eta k + \frac{1-\epsilon_2}{2}i\pi \quad {\rm mod}\, (2i\pi) \,,
\quad \epsilon_1 \epsilon_2 \epsilon_3=+1 \,, 
\label{constraint}
\ee
where $\epsilon_i=\pm 1$, and $k$ is an integer such that $|k|\leq
N-1$ and $N-1+k$ is even.  A further drawback of this solution is that
completeness is not straightforward, as two sets of Bethe Ansatz
equations are generally needed in order to obtain all $2^{N}$ levels
\cite{NR}.  Nevertheless, finite-size effects for this model and for
the boundary sine-Gordon model \cite{GZ} have been computed on the
basis of this solution \cite{finitesize}.  (Related results for the
boundary sine-Gordon model have been obtained by different methods in
\cite{TBA}.)  Many other interesting features, applications and
generalizations of this solution have also been found (see, e.g.,
\cite{Do}-\cite{Baj}).

There remains the vexing problem of solving the model when the
constraint (\ref{constraint}) is {\it not} satisfied, i.e., for {\it
arbitrary} values of the boundary parameters.  Our goal has been to
solve this problem for the root of unity case (\ref{etavalues}).  Some
progress was already achieved in \cite{MN1, MN2}, where two of us
(R.M. and R.N.) proposed Bethe-Ansatz-type solutions for special cases
with up to two free boundary parameters (and with the remaining
boundary parameters fixed to specific values).  For those special
cases (as well as for the cases where the constraint
(\ref{constraint}) is satisfied), the quantity $\Delta(u)$ defined by
\be 
\Delta(u) =  f(u)^{2} -4  \prod_{j=0}^{p} \delta(u+j\eta) \,,
\label{Delta}
\ee 
(where $f(u)$ and $\delta(u)$ are the functions appearing in the
functional relations, e.g. (\ref{p1functrltn}) - (\ref{p3functrltn}))
is a perfect square.  However, for generic values of boundary
parameters, the quantity $\Delta(u)$ is {\it not} a perfect square,
and it had not been clear to us how to proceed.  It is on this generic
case that we focus in this paper.

We find, for generic values of the boundary parameters, expressions
for the eigenvalues $\Lambda(u)$ of the transfer matrix $t(u)$ in
terms of sets of ``$Q$ functions'' $\{ a_{i}(u)\,, b_{i}(u) \}$, whose zeros
are given by Bethe-Ansatz-like equations.  (See (\ref{TQpgen3a}),
(\ref{TQpgen3b}) for $p>1$; and (\ref{TQp13a}), (\ref{TQp13b}) for
$p=1$.)  Such {\em generalized} $T-Q$ relations, involving more than one $Q$
function, appeared already for certain special cases \cite{MN2}, and
were used in \cite{MNS} to compute the corresponding boundary energies
in the thermodynamic limit.  We have verified the $T-Q$ relations
numerically for small values of $p$ and $N$, and confirmed that they
describe the complete set of $2^{N}$ eigenvalues.

The outline of this paper is as follows.  In Section
\ref{sec:transfer} we briefly review the transfer matrix and its
relation to the Hamiltonian (\ref{Hamiltonian}).  We present our
solution for the case $p > 1$ in Section \ref{sec:generalp}, and for
the XX chain ($p=1$) in Section \ref{sec:xx}.  We conclude in Section
\ref{sec:discuss} with a discussion of our results.  Some
details are relegated to the appendix.

\section{The transfer matrix}\label{sec:transfer}

The transfer matrix $t(u)$ of the model is given by \cite{Sk}
\be
t(u) = \tr_{0} K^{+}_{0}(u)\  
T_{0}(u)\  K^{-}_{0}(u)\ \hat T_{0}(u)\,,
\label{transfer}
\ee
where $T_{0}(u)$ and $\hat T_{0}(u)$ are the monodromy matrices 
\be
T_{0}(u) = R_{0N}(u) \cdots  R_{01}(u) \,,  \qquad 
\hat T_{0}(u) = R_{01}(u) \cdots  R_{0N}(u) \,,
\label{monodromy}
\ee
and $\tr_{0}$ denotes trace over the ``auxiliary space'' 0.
The $R$ matrix is given by
\be
R(u) = \left( \begin{array}{cccc}
	\sinh  (u + \eta) &0            &0           &0            \\
	0                 &\sinh  u     &\sinh \eta  &0            \\
	0                 &\sinh \eta   &\sinh  u    &0            \\
	0                 &0            &0           &\sinh  (u + \eta)
\end{array} \right) \,,
\label{bulkRmatrix}
\ee 
where $\eta$ is the bulk anisotropy parameter; and $K^{\mp}(u)$ are
$2 \times 2$ matrices whose components
are given by \cite{dVGR, GZ}
\be
K_{11}^{-}(u) &=& 2 \left( \sinh \alpha_{-} \cosh \beta_{-} \cosh u +
\cosh \alpha_{-} \sinh \beta_{-} \sinh u \right) \non \\
K_{22}^{-}(u) &=& 2 \left( \sinh \alpha_{-} \cosh \beta_{-} \cosh u -
\cosh \alpha_{-} \sinh \beta_{-} \sinh u \right) \non \\
K_{12}^{-}(u) &=& e^{\theta_{-}} \sinh  2u \,, \qquad 
K_{21}^{-}(u) = e^{-\theta_{-}} \sinh  2u \,,
\label{Kminuscomponents}
\ee
and
\be
K^{+}(u) = \left. K^{-}(-u-\eta)\right\vert_{(\alpha_-,\beta_-,\theta_-)\rightarrow
(-\alpha_+,-\beta_+,\theta_+)} \,,
\ee 
where $\alpha_{\mp} \,, \beta_{\mp} \,, \theta_{\mp}$ are the boundary
parameters.  \footnote{We use a
parametrization of the boundary parameters which differs from that in 
\cite{dVGR, GZ, Ne1}. Specifically, the matrices $K^{\mp}(u)$ are equal to those 
appearing in the second reference in \cite{Ne2} divided by the factors 
$\kappa_{\mp}$, respectively.}  

For $u=0$, the transfer matrix is given by
\be
t(0) = c_{0} \id  \,, \qquad
c_{0} =  -8 \sh^{2N}\eta \ch \eta \sh \alpha_{-} \sh \alpha_{+} 
\ch \beta_{-} \ch \beta_{+} \,.
\label{initial}
\ee
For $\eta \ne i\pi/2$, the Hamiltonian (\ref{Hamiltonian}) is related 
to the first derivative of the transfer matrix at $u=0$,
\be
\hh = c_{1} t'(0) + c_{2} \id \,,
\label{Htrelation}
\ee
where
\be
c_{1} &=& -\left( 16 \sinh^{2N-1} \eta \cosh \eta
\sinh \alpha_{-} \sinh \alpha_{+}
\cosh \beta_{-} \cosh \beta_{+} \right)^{-1} \,, \non \\
c_{2} &=& - {\sinh^{2}\eta  + N \cosh^{2}\eta\over 2 \cosh \eta} 
\label{c1c2}
\,.
\ee 
For the special case $\eta = i\pi/2$ (i.e., $p=1$),
\be
t(0)=0\,, \qquad t'(0) = d_{0} \id \,, \qquad 
d_{0} = (-1)^{N+1} 8 i \sh \alpha_{-} \sh \alpha_{+} 
\ch \beta_{-} \ch \beta_{+} \,, 
\label{initialp1}
\ee
and the Hamiltonian (\ref{Hamiltonian}) is related 
to the second derivative of the transfer matrix at $u=0$ \cite{Ne3},
\be
\hh = d_{1} t''(0) \,, \qquad
d_{1} = (-1)^{N+1} \left( 32 \sh \alpha_{-} \sh \alpha_{+} 
\ch \beta_{-} \ch \beta_{+} \right)^{-1} \,.
\label{Htrelationp1}
\ee

In addition to the fundamental commutativity property
\be
\left[ t(u)\,, t(v) \right] = 0  \,,
\label{commutativity}
\ee 
the transfer matrix also has $i \pi$ periodicity
\be
t(u+ i \pi) = t(u) \,,
\label{periodicity}
\ee
crossing symmetry
\be
t(-u - \eta)= t(u) \,,
\label{transfercrossing}
\ee
and the asymptotic behavior 
\be
t(u) \sim -\cosh(\theta_{-}-\theta_{+})
{e^{u(2N+4)+\eta (N+2)}\over 2^{2N+1}} \id + 
\ldots \qquad \mbox{for} \qquad
u\rightarrow \infty \,.
\label{transfasympt}
\ee

\section{The case $p > 1$}\label{sec:generalp}

We treat in this section the case (\ref{etavalues}) with
$p > 1$, i.e., bulk anisotropy values $\eta = {i \pi\over 3}\,, {i
\pi\over 4}\,, \ldots$.  Following Bazhanov and Reshetikhin \cite{BR},
we first recast the functional relations for the transfer matrix
eigenvalues $\Lambda(u)$ as the condition that a matrix $\M(u)$ have
zero determinant.  The equations for the corresponding null
eigenvector, together with a key Ansatz (\ref{Ansatzpgen1})-(\ref{Ansatzpgen2}), then
lead to the desired set of generalized $T-Q$ relations for
$\Lambda(u)$ (\ref{TQpgen3a}), (\ref{TQpgen3b}) and the associated
Bethe-Ansatz equations (\ref{BAEpgena})-(\ref{BAEnorm}).

\subsection{The matrix $\M(u)$}\label{subsec:M}

Our objective is to determine the eigenvalues $\Lambda(u)$ of the
transfer matrix $t(u)$.  As noted in the Introduction, the transfer
matrix satisfies a functional relation (e.g.,
(\ref{p1functrltn})-(\ref{p3functrltn})), where the functions
$\delta(u)$ and $f(u)$ are given in Appendix \ref{sec:expressions}.
By virtue of the commutativity property
(\ref{commutativity}), the eigenvalues satisfy the same functional
relation as the corresponding transfer matrix, as well as the
properties (\ref{periodicity}) - (\ref{transfasympt}). Hence, for example,
for $p=2$ the eigenvalues satisfy
\be
\Lambda(u)\ \Lambda(u+\eta)\ \Lambda(u+2\eta)  
&-&  \delta(u)\ \Lambda(u +2\eta)
- \delta(u+\eta)\ \Lambda(u)  \non \\
&-&  \delta(u+2\eta)\ \Lambda(u +\eta)  = f(u) \,.
\ee 

The first main step is to reformulate the functional relation as the
condition that the determinant of some matrix vanish.  To this end,
let us consider the $(p+1) \times (p+1)$ matrix $\M(u)$ given by
\be
\M(u) = \left(
\begin{array}{ccccccccc}
    \Lambda(u) & -m_{1}(u) & 0  & \ldots  & 0 & 0 & -n_{p+1}(u)  \\
    -n_{1}(u) & \Lambda(u+\eta) & -m_{2}(u)  & \ldots & 0 & 0 & 0  \\
    \vdots  & \vdots & \vdots & \ddots 
    & \vdots  & \vdots  & \vdots   \\
      0 & 0 & 0 & \ldots  & -n_{p-1}(u) &
       \Lambda(u+(p-1) \eta) & -m_{p}(u) \\
      -m_{p+1}(u)  & 0 & 0 & \ldots & 0 & -n_{p}(u) &
    \Lambda(u+p \eta)
\end{array} \right)  
\label{calM}
\ee
where the  matrix elements $\{ m_{j}(u) \,,  n_{j}(u) \}$ are still to
be determined. Evidently, this matrix is essentially tridiagonal, with 
nonzero elements also in the lower left and upper right corners.
One can verify that in order to recast the functional relations as 
\be
\det \M(u) = 0 \,,
\label{det}
\ee
it is sufficient that the off-diagonal matrix elements $\{ m_{j}(u)
\,, n_{j}(u) \}$ be periodic functions of $u$ with period $i \pi$, and
satisfy the conditions
\be
m_{j}(u)\  n_{j}(u) &=&\delta(u+(j-1)\eta) \,, \qquad j = 1\,, 2 \,, \ldots
\,, p+1 \,, \label{cond1} \\
\prod_{j=1}^{p+1} m_{j}(u) + \prod_{j=1}^{p+1} n_{j}(u) &=& f(u) \,.
\label{cond2}
\ee 

We now proceed to determine a set of off-diagonal matrix elements $\{
m_{j}(u) \,, n_{j}(u) \}$ which satisfies these conditions.  Using
(\ref{cond1}) to express $n_{j}(u)$ in terms of $m_{j}(u)$, and then
substituting into (\ref{cond2}), we immediately
see that the quantity $z(u) \equiv \prod_{j=1}^{p+1} m_{j}(u)$ must satisfy
\be
z(u) + {1\over z(u)}\prod_{j=0}^{p} \delta(u+ j\eta) = f(u) \,.
\ee 
This being a quadratic equation for $z(u)$, we readily obtain the two
solutions
\be
z^{\pm}(u)={1\over 2}\left(f(u) \pm \sqrt{\Delta(u)} \right) \,,
\label{zpmfirst}
\ee
where the discriminant $\Delta(u)$ is the quantity (\ref{Delta}) mentioned 
in the Introduction, 
\be 
\Delta(u) =  f(u)^{2} -4  \prod_{j=0}^{p} \delta(u+j\eta) \,.
\label{againDelta}
\ee 
In short, we must find a set of matrix elements $\{ m_{j}(u) \,,
n_{j}(u) \}$ which satisfies (\ref{cond1}) and also
\be
\prod_{j=1}^{p+1} m_{j}(u) = z^{\pm}(u) \,, \label{cond2more}
\ee 
where $z^{\pm}(u)$ is given by (\ref{zpmfirst}).

In previous work \cite{Ne2, MN1, MN2} we considered special cases for
which $\Delta(u)$ is a perfect square.  However, for generic values of
the boundary parameters, $\Delta(u)$ is {\it not} a perfect square.
Hence, the off-diagonal matrix elements {\it cannot} all be meromorphic
functions of $u$.

In order to determine these matrix elements, it is convenient to
recast the expression for $z^{\pm}(u)$ into a more manageable form.
Noting that (see (\ref{delta0}), (\ref{f0}) and (\ref{f0odd}))
\be
\prod_{j=0}^{p} \delta_{0}(u+j\eta) = f_{0}(u)^{2} \,,
\ee
we see that
\be
\Delta(u) = f_{0}(u)^{2}\ \Delta_{1}(u) \,, \label{Deltaid}
\ee
where we have defined
\be
\Delta_{1}(u) =  f_{1}(u)^{2} -4  \prod_{j=0}^{p} \delta_{1}(u+j\eta) \,.
\label{Delta1}
\ee 
It follows from (\ref{zpmfirst}) and (\ref{Deltaid})  that
\be
z^{\pm}(u) = f_{0}(u)\ z^{\pm}_{1}(u) \,,
\label{zpmsecond}
\ee 
where
\be
z^{\pm}_{1}(u)={1\over 2}\left(f_{1}(u) \pm \sqrt{\Delta_{1}(u)} \right) \,.
\label{zpm1first}
\ee
Using the explicit expressions for $\delta_{1}(u)$ (\ref{delta1}) and $f_{1}(u)$ 
(\ref{f1even}), (\ref{f1odd}), one can show that $\Delta_{1}(u)$ (\ref{Delta1}) can be
expressed as 
\be
\Delta_{1}(u) = 4 \sinh^{2}(2(p+1)u) \sum_{k=0}^{2} \mu_{k} \cosh^{k}(2(p+1)u) \,,
\label{Delta1explicit}
\ee 
where the coefficients $\mu_{k}$, which depend on the boundary
parameters, are given in the Appendix (\ref{mueven}), (\ref{muodd})
for even and odd values of $p$, respectively.  
It follows from (\ref{zpm1first}) and (\ref{Delta1explicit}) that 
\be
z^{\pm}_{1}(u)={1\over 2}\left(f_{1}(u) \pm g_{1}(u)\ Y(u) \right) \,,
\label{zpm1second}
\ee
where we have defined
\be
g_{1}(u) = 2 \sinh (2(p+1)u) 
\label{g}
\ee
and 
\be 
Y(u) = \sqrt{\sum_{k=0}^{2} \mu_{k} \cosh^{k}(2(p+1)u)} \,,
\label{Y}
\ee 
which we take to be a single-valued continuous branch obtained 
by introducing suitable branch cuts in the complex 
$u$ plane.\footnote{We assume that
the boundary parameters have generic values, and therefore, the
function $\sum_{k=0}^{2} \mu_{k} \cosh^{k}(2(p+1)u)$ is not a perfect square. 
The branch points are zeros of this function.}
One can see that $Y(u)$ has the properties 
\be
Y(u+\eta) = Y(u) \,, \qquad Y(-u) = Y(u) \,.
\label{Yprops}
\ee 
It follows from (\ref{fident}), (\ref{zpm1second}) and (\ref{g}) that 
\be
z^{\pm}_{1}(u+\eta) = z^{\pm}_{1}(u) \qquad z^{+}_{1}(-u) = z^{-}_{1}(u)
\,. \label{zpm1props}
\ee
In short, $z^{\pm}(u)$ is given by (\ref{zpmsecond}), where  
$z^{\pm}_{1}(u)$ is given by (\ref{zpm1second}) - (\ref{Y}), and
has the important properties (\ref{zpm1props}).

In order to construct the desired set of matrix elements, it is also
convenient to introduce the function $h(u)$,
\be
h(u) = h_{0}(u)\ h_{1}(u) \,,
\label{hfunction}
\ee
where $h_{0}(u)$ is given by
\be
h_{0}(u) = (-1)^{N}\sinh^{2N}(u+\eta){\sinh(2u+2\eta)\over \sinh(2u+\eta)} \,,
\label{h0}
\ee 
and satisfies
\be
h_{0}(u)\ h_{0}(-u) &=& \delta_{0}(u-\eta) \,, \label{h0prop1} \\
\prod_{k=0}^{p} h_{0}(u+ k\eta) = \prod_{k=0}^{p} h_{0}(-u-k\eta) &=&
f_{0}(u) \,. \label{h0prop2}
\ee
Moreover, $h_{1}(u)$ is given by \footnote{Presumably, one can use 
the more general expression 
$h_{1}(u) = (-1)^{N+1} 4 \sinh(u+\alpha_{-}) \cosh(u+ \epsilon_{1}\beta_{-}) 
\sinh(u+ \epsilon_{2}\alpha_{+}) \cosh(u+ \epsilon_{3}\beta_{+})$, 
where $\epsilon_{i}= \pm 1$, which also satisfies (\ref{h1b}). 
However, for simplicity, we restrict to the special case $\epsilon_{i}= 1$.}
\be
h_{1}(u) = (-1)^{N+1} 4 \sinh(u+\alpha_{-}) \cosh(u+\beta_{-}) 
\sinh(u+\alpha_{+}) \cosh(u+\beta_{+}) \,,
\label{h1a}
\ee 
and satisfies
\be
h_{1}(u)\ h_{1}(-u) = \delta_{1}(u-\eta) \,. \label{h1b} 
\ee

We are finally ready to explicitly construct the requisite matrix 
elements:
\be
m_{j}(u) &=& h(-u-j \eta) \,, \qquad  n_{j}(u) = h(u+j \eta) \,, \qquad
j=1\,, 2\,, \ldots \,, p \,, \non \\
m_{p+1}(u) &=& {z^{-}(u) \over 
\prod_{k=1}^{p}h(-u-k\eta)} = {z^{-}_{1}(u)\ h_{0}(-u) \over 
\prod_{k=1}^{p}h_{1}(-u-k\eta)} \,, \non \\
n_{p+1}(u) &=& {z^{+}(u) \over 
\prod_{k=1}^{p}h(u+k\eta)} = {z^{+}_{1}(u)\ h_{0}(u) \over 
\prod_{k=1}^{p}h_{1}(u+k\eta)} \,,
\label{matelements} 
\ee 
Indeed, using (\ref{h0prop1}), (\ref{h1b}) and the fact
\be
z^{+}(u)\ z^{-}(u) = \prod_{j=0}^{p} \delta(u+j\eta) 
\ee
(which follows from (\ref{zpmfirst}) and (\ref{againDelta})), it is easy 
to see that the condition (\ref{cond1}) is satisfied. It is also easy 
to see that the condition (\ref{cond2more}) (with $z^{-}(u)$ on the RHS)
is also satisfied. We note here for future reference that 
\be
n_{p+1}(u) = m_{p+1}(-u) \,, 
\label{melementscrossing}
\ee
which follows from (\ref{zpm1props}). 
We also note that if the constraint (\ref{constraint}) with
$\epsilon_{1}=\epsilon_{2}=\epsilon_{3}=1$ is satisfied, then
$\prod_{k=0}^{p}h_{1}(u+k\eta) = z_{1}^{\pm}(u)$, as follows from the
identity (A.8) in \cite{MN1}.  Hence, for this case, $n_{p+1}(u) =
h(u)$, and the matrix $\M(u)$ reduces to the one considered in
\cite{Ne2}.

\subsection{Bethe Ansatz}\label{subsec:BA}

The fact (\ref{det}) that $\M(u)$ has a zero determinant implies that it 
has a null eigenvector 
$v(u) = (v_{1}(u) \,, v_{2}(u)\,, \ldots  \,, v_{p+1}(u))$,
\be
\M(u)\ v(u) = 0 \,. \label{nulleigenvector1}
\ee 
We shall assume the periodicity 
\be
v_{j}(u + i\pi) = v_{j}(u+ (p+1)\eta) = v_{j}(u) \,,
\qquad j = 1\,, \ldots \,, p+1 \,,
\label{vperiodicity}
\ee
which is consistent with the periodicity $\M(u + i\pi) = \M(u)$.
It follows from (\ref{nulleigenvector1}) and the 
expression (\ref{calM}) for $\M(u)$ that
\be
\Lambda(u+(j-1)\eta)\ v_{j}(u) = n_{j-1}(u)\ v_{j-1}(u) + m_{j}(u)\
v_{j+1}(u) \,, \quad  j = 1\,, 2 \,, \ldots \,, p+1 \,,
\ee
where $v_{j+p+1} = v_{j}$ and $n_{j+p+1} = n_{j}$. Shifting $u \mapsto
u - (j-1)\eta$, we readily obtain
\be
\Lambda(u)\ v_{1}(u) &=& h(-u-\eta)\ v_{2}(u) + n_{p+1}(u)\ v_{p+1}(u) \,,
\non \\
\Lambda(u)\ v_{j}(u-(j-1)\eta) &=& h(u)\ v_{j-1}(u-(j-1)\eta) +
h(-u-\eta)\ v_{j+1}(u-(j-1)\eta) \,, \non \\
& & \qquad  \quad  j = 2\,, 3\,, \ldots \,, p \,, \non \\ 
\Lambda(u)\ v_{p+1}(u-p\eta) &=& h(u)\ v_{p}(u-p\eta) +
m_{p+1}(u-p\eta)\ v_{1}(u-p\eta) \,.
\label{TQpgen1}
\ee
The crossing properties of the eigenvalue $\Lambda(-u - \eta)= \Lambda(u)$
(\ref{transfercrossing}) together with (\ref{melementscrossing})
suggest a corresponding crossing property of $v(u)$, namely,
\be
v_{j}(-u) = v_{p+2-j}(u) \,, \qquad j = 1\,, 2\,, \ldots \,, p+1 \,.
\label{vcrossing}
\ee 
In particular, for $j={p\over 2}+1$ (which occurs only for $p$ even !), 
this relation implies that $v_{{p\over 2}+1}(u)$ is crossing
invariant,
\be
v_{{p\over 2}+1}(-u) = v_{{p\over 2}+1}(u) \,.
\label{v1crossing}
\ee 
Moreover, (\ref{vcrossing}) implies that at most $\lfloor{p\over
2}\rfloor+1$ components of $v(u)$ are independent, say, $\{
v_{1}(u)\,, \ldots$ $\,, v_{\lfloor{p\over 2}\rfloor+1}(u) \}$, where
$\lfloor\quad \rfloor$ denotes integer part.

Substituting the explicit expression for $n_{p+1}(u)$
(\ref{matelements}) into (\ref{TQpgen1}), we obtain the relations
\be
\Lambda(u)\ v_{1}(u) &=& h(-u-\eta)\ v_{2}(u) + {z^{+}_{1}(u)\ h_{0}(u) \over 
\prod_{k=1}^{p}h_{1}(u+k\eta)}\ v_{1}(-u)  \,, \non \\ 
\Lambda(u)\ v_{j}(u-(j-1)\eta) &=& h(u)\ v_{j-1}(u-(j-1)\eta) +
h(-u-\eta)\ v_{j+1}(u-(j-1)\eta) \,, \non \\
& & \qquad  \quad    j = 2 \,, \ldots \,,
\lfloor{p\over 2}\rfloor + 1 \,,
\label{TQpgen2}
\ee
which evidently resemble a system of generalized $T-Q$ equations. 
However, since $\Lambda(u)$ is an analytic function of $u$ for finite 
values of $u$ \footnote{This is a well-known consequence of the transfer matrix 
properties (\ref{commutativity}) - (\ref{transfasympt}).}, the functions $v_{j}(u)$ 
{\it cannot} be analytic due to the presence of the $z^{+}_{1}(u)$
factor in (\ref{TQpgen2}).

We therefore propose instead the following Ansatz:
\be
v_{j}(u) = a_{j}(u) + b_{j}(u)\ Y(u) \,, \qquad j = 1\,, 2 \,, \ldots 
\,, \lfloor{p\over 2}\rfloor +1 \,, 
\label{Ansatzpgen1}
\ee
where $Y(u)$ is the function (\ref{Y}), and $a_{j}(u)\,, b_{j}(u)$ are
periodic, analytic functions of $u$,
\be
a_{j}(u) &=& A_{j} \prod_{k=1}^{2M_{a}} \sinh(u-u_{k}^{(a_{j})}) \,, \qquad 
b_{j}(u) = B_{j}\prod_{k=1}^{2M_{b}} \sinh(u-u_{k}^{(b_{j})})\,,
\qquad j \ne {p\over 2}+1 \,, \non \\
a_{{p\over 2}+1}(u) &=& A_{{p\over 2}+1} 
\prod_{k=1}^{M_{a}} \sinh(u-u_{k}^{(a_{{p\over 2}+1})})
\sinh(u+u_{k}^{(a_{{p\over 2}+1})}) \,, \non \\
b_{{p\over 2}+1}(u) &=& B_{{p\over 2}+1}
\prod_{k=1}^{M_{b}} \sinh(u-u_{k}^{(b_{{p\over 2}+1})}) 
\sinh(u+u_{k}^{(b_{{p\over 2}+1})}) \,, 
\label{Ansatzpgen2}
\ee
whose zeros $\{ u_{k}^{(a_{j})} \,, u_{k}^{(b_{j})} \}$, normalization
constants $\{ A_{j}\,, B_{j} \}$, and also the integers $M_{a}\,, M_{b}$ are still to be
determined. \footnote{Since the normalization of the null eigenvector $v(u)$ is arbitrary,
one of the normalization constants, say $B_{1}$, can be set to unity.}
The forms (\ref{Ansatzpgen2}) for $a_{j}(u)$ and
$b_{j}(u)$ evidently have the periodicity and crossing properties
\be
a_{j}(u+ i\pi) &=& a_{j}(u) \,, \qquad \quad b_{j}(u+ i\pi) = b_{j}(u) \,, 
\qquad j = 1\,, \ldots \,, p+1 \,, \non \\
a_{{p\over 2}+1}(-u) &=& a_{{p\over 2}+1}(u) \,, \qquad
b_{{p\over 2}+1}(-u) = b_{{p\over 2}+1}(u) \,,
\ee 
which reflect the corresponding properties of $v_{j}(u)$
(\ref{vperiodicity}), (\ref{v1crossing}) and of $Y(u)$ (\ref{Yprops}).
We have obtained numerical support for this Ansatz, which we discuss
at the end of this section.

We now substitute the Ansatz (\ref{Ansatzpgen1}), as well as the
expression for $z^{+}_{1}(u)$ (\ref{zpm1second}), into 
(\ref{TQpgen2}).
Since $\Lambda(u)$ and $Y(u)^{2}$ (but not $Y(u)$ !)  are analytic
function of $u$, we can separately equate the terms that are linear in
$Y(u)$, and the terms with even (i.e., $0$ or $2$) powers of $Y(u)$.
In this way we finally arrive at the generalized $T-Q$ equations:
\be
\Lambda(u)\ a_{1}(u)\ 
= h(-u-\eta)\ a_{2}(u) 
+{h_{0}(u) \over 
2\prod_{k=1}^{p}h_{1}(u+k\eta)}
\Big[f_{1}(u)\ a_{1}(-u) + g_{1}(u)\ Y(u)^{2}\ b_{1}(-u) \Big] \,, 
\non 
\ee
\be
\Lambda(u)\ a_{j}(u-(j-1)\eta) &=& h(u)\ a_{j-1}(u-(j-1)\eta) +
h(-u-\eta)\ a_{j+1}(u-(j-1)\eta) \,, \non \\
& & \qquad \qquad j = 2 \,, \ldots 
\,, \lfloor{p\over 2}\rfloor +1 \,, 
\label{TQpgen3a}
\ee
and
\be
\Lambda(u)\ b_{1}(u)\ 
= h(-u-\eta)\ b_{2}(u) 
+{h_{0}(u) \over 
2\prod_{k=1}^{p}h_{1}(u+k\eta)}
\Big[f_{1}(u)\ b_{1}(-u) + g_{1}(u)\ a_{1}(-u) \Big] \,, 
\non 
\ee
\be
\Lambda(u)\ b_{j}(u-(j-1)\eta) &=& h(u)\ b_{j-1}(u-(j-1)\eta) +
h(-u-\eta)\ b_{j+1}(u-(j-1)\eta) \,, \non \\
& & \qquad \qquad j = 2 \,, \ldots 
\,, \lfloor{p\over 2}\rfloor +1 \,, 
\label{TQpgen3b}
\ee
where $a_{{p\over2}+2}(u)=a_{{p\over2}}(-u)$ and
$a_{{p+3\over2}}(u)=a_{{p+1\over2}}(-u)$ for even and odd values of
$p$, respectively, 
and similarly for the $b$'s. 

The asymptotic behavior $\Lambda(u) \sim e^{u(2N+4)}$ for
$u\rightarrow \infty$ (\ref{transfasympt}) together with the $T-Q$
equations imply the relation
\be
M_{a} = M_{b} + p + 1 \,.
\label{mambrelation}
\ee 
An analysis of the $u$-independent terms yields relations among the 
normalization constants and sums of zeros ($\sum_{l} u_{l}^{(a_{j})}\,,
\sum_{l} u_{l}^{(b_{j})}$), which we do not record here.

As usual, analyticity of $\Lambda(u)$ and the $T-Q$ equations imply
Bethe-Ansatz-like equations for the zeros $\{ u_{l}^{(a_{j})} \}$ of
the functions $\{ a_{j}(u)\}$,
\be
{h_{0}(-u_{l}^{(a_{1})}-\eta)\over h_{0}(u_{l}^{(a_{1})})}
&=&-{f_{1}(u_{l}^{(a_{1})})\ a_{1}(-u_{l}^{(a_{1})}) +
g_{1}(u_{l}^{(a_{1})})\ Y(u_{l}^{(a_{1})})^{2}\ 
b_{1}(-u_{l}^{(a_{1})})\over
2a_{2}(u_{l}^{(a_{1})})\ h_{1}(-u_{l}^{(a_{1})}-\eta)\
\prod_{k=1}^{p}h_{1}(u_{l}^{(a_{1})}+k\eta)} \,, \non \\
{h(-u_{l}^{(a_{j})}-j\eta)\over 
h(u_{l}^{(a_{j})}+(j-1)\eta)}&=&-{a_{j-1}(u_{l}^{(a_{j})})\over
a_{j+1}(u_{l}^{(a_{j})})}  \,, \qquad 
j = 2 \,, \ldots \,, \lfloor{p\over 2}\rfloor+1 \,,
\label{BAEpgena}
\ee 
and for the zeros $\{ u_{l}^{(b_{j})} \}$ of the functions $\{ b_{j}(u)\}$,
\be
{h_{0}(-u_{l}^{(b_{1})}-\eta)\over h_{0}(u_{l}^{(b_{1})})}
&=&-{f_{1}(u_{l}^{(b_{1})})\ b_{1}(-u_{l}^{(b_{1})}) +
g_{1}(u_{l}^{(b_{1})})\  
a_{1}(-u_{l}^{(b_{1})})\over
2b_{2}(u_{l}^{(b_{1})})\ h_{1}(-u_{l}^{(b_{1})}-\eta)\
\prod_{k=1}^{p}h_{1}(u_{l}^{(b_{1})}+k\eta)} \,, \non \\
{h(-u_{l}^{(b_{j})}-j\eta)\over 
h(u_{l}^{(b_{j})}+(j-1)\eta)}&=&-{b_{j-1}(u_{l}^{(b_{j})})\over
b_{j+1}(u_{l}^{(b_{j})})}  \,, \qquad 
j = 2 \,, \ldots \,, \lfloor{p\over 2}\rfloor+1 \,.
\label{BAEpgenb}
\ee 
Moreover, there are additional Bethe-Ansatz-like equations for the
normalization constants. Indeed, noting that $h_{0}(u)$ has a
pole at $u=-{\eta\over 2}$, it follows from the analyticity of $\Lambda(u)$ and 
the $T-Q$ equations (\ref{TQpgen3a}) that
\be
a_{1}({\eta\over 2}) &=& a_{2}(-{\eta\over 2}) \,,
\label{newbaea1} \\
a_{j-1}(({1\over 2}-j)\eta) &=&  a_{j+1}(({1\over 2}-j)\eta) \,, 
\qquad j = 2 \,, \ldots \,, \lfloor{p\over 2}\rfloor +1 \,.
\label{newbaeaj}
\ee 
In obtaining the first equation (\ref{newbaea1}), we have made use of 
the identity
\be
f_{1}(-{\eta\over 2}) =  2 \prod_{k=0}^{p} h_{1}(-{\eta\over 2}
+ \eta k) \,.
\ee
The equations (\ref{newbaea1}), (\ref{newbaeaj}) evidently further
relate the normalization constants $\{ A_{j} \}$. Similarly, the
$T-Q$ equations (\ref{TQpgen3b}) imply
\be
b_{1}({\eta\over 2}) &=& b_{2}(-{\eta\over 2}) \,,
\label{newbaeb1}\\
b_{j-1}(({1\over 2}-j)\eta) &=&  b_{j+1}(({1\over 2}-j)\eta) \,, 
\qquad j = 2 \,, \ldots \,, \lfloor{p\over 2}\rfloor +1 \,, 
\label{newbaebj}
\ee 
which relate the normalization constants $\{ B_{j} \}$. Finally,
noting that the first (i.e., $j=1$) $T-Q$ equation in the set (\ref{TQpgen3b}) 
has the factor $\prod_{k=1}^{p}h_{1}(u+k\eta)$ in the denominator which can vanish, 
e.g. at $u=-\alpha_{-}-\eta$, leads to the relation 
\be
f_{1}(-\alpha_{-}-\eta)\ b_{1}(\alpha_{-}+\eta) =
-g_{1}(-\alpha_{-}-\eta)\  a_{1}(\alpha_{-}+\eta)
\,,
\label{BAEnorm}
\ee
which relates the normalization constants $A_{1}$ and $B_{1}$. A
similar analysis of the first equation in (\ref{TQpgen3a}) gives an
equivalent result, by virtue of the identity $f_{1}(u_{0})^{2} =
g_{1}(u_{0})^{2} Y(u_{0})^{2}$ if $u_{0}$ satisfies
$\prod_{j=0}^{p} \delta_{1}(u_{0}+j\eta) 
=0$, which follows from (\ref{Delta1}) and the fact (\ref{Delta1explicit}) that
$\Delta_{1}(u) = g_{1}(u)^{2} Y(u)^{2}$.

The energy eigenvalues of the Hamiltonian (\ref{Hamiltonian}) follow 
from (\ref{initial})-(\ref{c1c2}) and the $T-Q$ relations 
(\ref{TQpgen3a}),
\be
E = c_{1} \Lambda'(0) + c_{2} =  c_{1} c_{0} 
\left[ -{a_{j}'(-(j-1)\eta)\over a_{j}(-(j-1)\eta)} + 
{a_{j-1}'(-(j-1)\eta)\over a_{j-1}(-(j-1)\eta)}
+ {h'(0)\over h(0)} \right] + c_{2}  \,,
\label{energyinitial}
\ee 
where $j$ can take any value in the set $\{2\,, \ldots \,, \lfloor{p\over 2}\rfloor+1\}$. For 
$j \ne {p\over 2}+1$, it follows that 
\be
E &=& {1\over 2}\sinh \eta \sum_{l=1}^{2M_{a}} \left[
\coth (u_{l}^{(a_{j})}+(j-1)\eta) - \coth (u_{l}^{(a_{j-1})}+(j-1)\eta) \right]
\non \\
&+& {1\over 2}\sinh \eta \left( \coth \alpha_{-} + \tanh \beta_{-} 
+ \coth \alpha_{+} + \tanh \beta_{+} \right) 
+ {1\over 2}(N-1)\cosh \eta\,, \label{energypgen}
\ee
which does not depend explicitly on the normalization constants.
For $j = {p\over 2}+1$, there is an additional 
contribution from the term ${a_{j}'(-(j-1)\eta)\over a_{j}(-(j-1)\eta)}$ 
in (\ref{energyinitial}), 
since this $a_{j}(u)$ is crossing invariant. It follows that 
\be
E= {1\over 2}\sinh \eta  \left\{ \sum_{l=1}^{M_{a}} \left[
\coth (u_{l}^{(a_{{p\over 2}+1})}+{p\eta\over 2}) - 
\coth (u_{l}^{(a_{{p\over 2}+1})}-{p\eta\over 2}) \right]
- \sum_{l=1}^{2M_{a}} \coth (u_{l}^{(a_{p\over 2})}+{p\eta\over 2}) \right\}  + \ldots 
\non \\
\label{energyspecial}
\ee 
where the ellipsis denotes the terms in (\ref{energypgen}) that are independent of Bethe roots.
If one works instead with the $T-Q$ relations (\ref{TQpgen3b}), one
obtains the same results (\ref{energyspecial}), (\ref{energypgen}),
except with sums over the $b$ roots.

We have verified the $T-Q$ equations numerically, for values of $p$
and $N$ up to $6$ and for generic values of the boundary parameters, along
the lines \cite{NR}. These results are consistent with the conjecture
\be
M_{a} = \lfloor {N-1\over 2} \rfloor + 2p+1 \,, \qquad 
M_{b} = \lfloor {N-1\over 2} \rfloor + p \,,
\label{Mvaluespgen}
\ee
which agrees with the relation (\ref{mambrelation}).  These numerical
results also indicate that our Bethe Ansatz solution is complete: for
each value of $N$, we find sets of Bethe roots corresponding to each
of the $2^{N}$ eigenvalues of the transfer matrix.  As already
remarked, this numerical work provides support for the Ansatz
(\ref{Ansatzpgen1}), (\ref{Ansatzpgen2}).

\section{The XX chain $(p=1)$}\label{sec:xx}  

The case $p=1$ corresponds to bulk anisotropy value $\eta= i \pi/2$,
for which the bulk Hamiltonian (\ref{bulkHamiltonian}) reduces to 
\be 
\hh_{0} = {1\over 2}\sum_{n=1}^{N-1}\left( 
\sigma_{n}^{x}\sigma_{n+1}^{x}+\sigma_{n}^{y}\sigma_{n+1}^{y}
\right) \,,
\label{xx}
\ee
which is known as the XX chain.  The open XX chain with nondiagonal
boundary terms was studied earlier in \cite{Ne3}-\cite{BiRi}.

The functional equation for the case $p=1$ is given by (\ref{p1functrltn}).
We find that a suitable matrix $M(u)$ is given by
\be
\M(u) = \left(
\begin{array}{cc}
    \Lambda(u) & -m_{1}(u)   \\
   -n_{1}(u) & \Lambda(u+\eta)  \\
\end{array} \right)  \,,
\ee
where
\be
m_{1}(u) &=& {1\over h_{1}(-u)}\left[ h_{0}(u)\ \delta_{1}(u+\eta) +
h_{0}(-u-\eta)\ z_{1}^{-}(u) \right] \,, \label{moddp1} \\
n_{1}(u) &=& {1\over h_{1}(u)}\left[ h_{0}(-u)\ \delta_{1}(u+\eta) +
h_{0}(u+\eta)\ z_{1}^{+}(u) \right] \,.
\ee 
Indeed, one can verify that the condition $\det \M(u)=0$ reproduces
the functional equation (\ref{p1functrltn}). Note that
\be
n_{1}(u) = m_{1}(-u) \,.
\label{melementscrossingp1}
\ee
The corresponding null eigenvector $v(u) = (v_{1}(u) \,, v_{2}(u))$
satisfies $\M(u)\ v(u) = 0$, i.e.,
\be
\Lambda(u)\ v_{1}(u) &=& m_{1}(u)\ v_{2}(u) \,, \label{TQp11} \\
\Lambda(u)\ v_{2}(u-\eta) &=& n_{1}(u-\eta)\ v_{1}(u-\eta) \,.
\ee
The crossing symmetry $\Lambda(-u - \eta)= \Lambda(u)$ 
and (\ref{melementscrossingp1}) suggest 
\be
v_{2}(u) = v_{1}(-u)  \,.
\label{vcrossingp1}
\ee 
That is, only one component is independent, say, $v_{1}(u)$.
Substituting the explicit expression (\ref{moddp1}) into the first
equation (\ref{TQp11}), we obtain
\be
\Lambda(u)\ v_{1}(u) = {1\over h_{1}(-u)}\left[ h_{0}(u)\ \delta_{1}(u+\eta) +
h_{0}(-u-\eta)\ z_{1}^{-}(u) \right] v_{1}(-u) \,.
\label{TQp12}
\ee 
Similarly to the $p>1$ case, we make the Ansatz
\be
v_{1}(u) = a_{1}(u) + b_{1}(u)\ Y(u) \,, 
\label{Ansatzp1first}
\ee
where 
\be
a_{1}(u) = A \prod_{k=1}^{2M_{a}} \sinh(u-u_{k}^{(a_{1})}) \,, \qquad 
b_{1}(u) = \prod_{k=1}^{2M_{b}} \sinh(u-u_{k}^{(b_{1})})\,.
\label{Ansatzp1second}
\ee
Substituting this Ansatz, together with the
expression for $z^{-}_{1}(u)$ (\ref{zpm1second}), into 
(\ref{TQp12}), we obtain the desired generalized $T-Q$ equations:
\be
\Lambda(u)\ a_{1}(u)\ h_{1}(-u) &=& \left[ h_{0}(u)\ \delta_{1}(u+\eta) 
+{1\over 2} h_{0}(-u-\eta)\ f_{1}(u) \right] a_{1}(-u) \non \\
&-& {1\over 2} h_{0}(-u-\eta)\ g_{1}(u)\ Y(u)^{2}\ b_{1}(-u) \,,
\label{TQp13a} \\ 
\Lambda(u)\ b_{1}(u)\ h_{1}(-u) &=& \left[ h_{0}(u)\ \delta_{1}(u+\eta) 
+{1\over 2} h_{0}(-u-\eta)\ f_{1}(u) \right] b_{1}(-u) \non \\
&-& {1\over 2} h_{0}(-u-\eta)\ g_{1}(u)\ a_{1}(-u)  \,.
\label{TQp13b}
\ee 
From the asymptotic behavior we obtain the relation
\be
M_{a} = M_{b} + 2 \,.
\label{mambp1}
\ee 
The corresponding Bethe Ansatz equations are
\be
{h_{0}(-u_{l}^{(a_{1})}-\eta)\over h_{0}(u_{l}^{(a_{1})})}
&=&-{2\delta_{1}(u_{l}^{(a_{1})}+\eta)\ a_{1}(-u_{l}^{(a_{1})}) \over
f_{1}(u_{l}^{(a_{1})})\ a_{1}(-u_{l}^{(a_{1})}) -
g_{1}(u_{l}^{(a_{1})})\ Y(u_{l}^{(a_{1})})^{2}\ 
b_{1}(-u_{l}^{(a_{1})})} \,, \label{BAEp1a} \\
{h_{0}(-u_{l}^{(b_{1})}-\eta)\over h_{0}(u_{l}^{(b_{1})})}
&=&-{2\delta_{1}(u_{l}^{(b_{1})}+\eta)\ b_{1}(-u_{l}^{(b_{1})}) \over
f_{1}(u_{l}^{(b_{1})})\ b_{1}(-u_{l}^{(b_{1})}) -
g_{1}(u_{l}^{(b_{1})})\ 
a_{1}(-u_{l}^{(b_{1})})} \,, \label{BAEp1b}
\ee 
and, e.g., 
\be
f_{1}(\alpha_{-})\ b_{1}(-\alpha_{-}) =
g_{1}(\alpha_{-})\  a_{1}(-\alpha_{-})
\,.
\ee
There are no additional relations arising from analyticity at $u=-{\eta\over
2}$ analogous to (\ref{newbaea1}),  (\ref{newbaeb1}) due to the identity 
$f_{1}(-{\eta\over 2}) = 2 \delta_{1}({\eta\over 2})$.

The energy eigenvalues of the Hamiltonian (\ref{Hamiltonian}) follow 
from (\ref{initialp1}), (\ref{Htrelationp1}) and the first $T-Q$ relation 
(\ref{TQp13a}),
\be
E &=& d_{1} \Lambda''(0) = d_{1} d_{0} \left[ -2{a_{1}'(0)\over a_{1}(0)} 
+ {h_{1}'(0)\over h_{1}(0)} \right]  \non \\
&=& i \sum_{l=1}^{2M_{a}} \coth  u_{l}^{(a_{1})} + {i\over
2}\left( \coth \alpha_{-} + \tanh \beta_{-} + \coth \alpha_{+} + \tanh
\beta_{+} \right) \,. 
\label{energyp1}
\ee
Working instead with the second $T-Q$ relation (\ref{TQp13b}) gives 
the same result (\ref{energyp1}) except with sums over $b$ roots.

We have verified that the above $T-Q$ equations are well-satisfied
numerically, for values of $N$ up to $8$ and for generic values of the
boundary parameters, along the lines \cite{NR}.  These results are
consistent with the conjecture 
\be
M_{a} =  \lfloor {N-1\over 2} \rfloor + 3  \,, \qquad 
M_{b} =  \lfloor {N-1\over 2} \rfloor + 1 \,,
\label{Mvaluesp1}
\ee
which agrees with the relation (\ref{mambp1}), and in fact also with
(\ref{Mvaluespgen}).
These results also indicate that our Bethe Ansatz solution
is complete.

\section{Discussion}\label{sec:discuss}

We have found a Bethe-Ansatz-type solution of the open spin-$1/2$
integrable XXZ quantum spin chain with general integrable boundary
terms at roots of unity.  All six boundary parameters are arbitrary.
In particular, the boundary parameters need {\it not} satisfy the
constraint (\ref{constraint}) that arose in previous work
\cite{Ne2}-\cite{YNZ}.  Moreover, in contrast to that earlier
solution, our new solution appears to give the complete set of $2^{N}$
eigenvalues in a straightforward manner.  This solution is essentially
the same for both even and odd values of $p$, the main difference
being that, in the former case, one of the $Q$ functions is
crossing invariant.

Part of the price paid for this success is that there are multiple 
$Q$ functions $\{ a_{j}(u) \,, b_{j}(u) \}$ and corresponding 
multiple sets of Bethe roots $\{ u_{l}^{(a_{j})} \,, u_{l}^{(b_{j})} 
\}$. However, we have already demonstrated the feasibility of
performing thermodynamic ($N \rightarrow \infty$) computations
with two such sets of Bethe roots \cite{MNS}. Hence, we expect 
that this multiplicity of sets of Bethe roots will not cause 
significant computational difficulty. A further complication is the
appearance of normalization constants $\{ A_{j}\,, B_{j} \}$ and 
their corresponding Bethe-Ansatz-type equations.

Another part of the price paid for this success is that the bulk
anisotropy parameter is restricted to the values (\ref{etavalues}).
However, we expect that it should be possible to further generalize
our solution to the case $\eta = i \pi p'/(p+1)$, where $p'$ is also
an integer.  Indeed, we expect that functional relations of order
$p+1$ with the same structure (e.g.,
(\ref{p1functrltn})-(\ref{p3functrltn})) will continue to hold for
that case, except with a different function $f(u)$ that will now
depend also on $p'$.  Hence, to the extent that a number can be
approximated by a rational number, this approach should in principle
solve the problem for general imaginary values of $\eta$.
Unfortunately, this approach does not seem to be suitable for directly
addressing the problem of real values of $\eta$, for which case the
transfer matrix presumably does not obey functional relations of
finite order.  Nevertheless, as in the case of the sinh-Gordon and
sine-Gordon models, it may perhaps be possible to obtain results for
real values of $\eta$ from those of imaginary values of $\eta$ by some
sort of analytic continuation.

Although we have considered here the case of generic values of the
boundary parameters for which the quantity $\Delta(u)$ (\ref{Delta})
is not a perfect square, we find numerical evidence that our solution
remains valid when $\Delta(u)$ becomes a perfect square.  Presumably,
for such special cases, the $Q$ functions $\{ a_{j}(u) \,, b_{j}(u)
\}$ are not independent.  It may be interesting to determine the
precise relationship between these $Q$ functions and those appearing
in the previously found solutions \cite{Ne2}-\cite{YNZ}, \cite{MN1,
MN2}.

We remark that the set of off-diagonal elements (\ref{matelements}) of
the matrix $\M(u)$ is not unique.  Indeed, we have found other sets of
matrix elements which also give $\det \M(u) = 0$.  Among all the sets
which we found, the particular set presented here has several
advantages: (i) it works for both even and odd values of $p$; (ii) the
corresponding $T-Q$ relations and Bethe Ansatz equations are
relatively simple; (iii) the corresponding values of $M_{a}$ and
$M_{b}$ are minimized.  Nevertheless, it may be worthwhile to continue
looking for alternative sets of off-diagonal matrix elements, which
may further reduce the values of $M_{a}$ and $M_{b}$, or which may
have other nice properties.
 
A key step in our analysis is the Ansatz (\ref{Ansatzpgen1}),
(\ref{Ansatzpgen2}), which allows us to express the non-analytic
quantities $\{ v_{j}(u) \}$ in terms of analytic ones $\{ a_{j}(u)
\,, b_{j}(u) \}$.  We have numerical evidence that this Ansatz is
valid.  However, it is not clear whether this Ansatz is the most
``economical'': there may be alternative Ans\"atze which introduce
fewer $Q$ functions.  For example, there may be some fixed relation
between $a_{j}(u)$ and $b_{j}(u)$.

The structure of our generalized $T-Q$ equations bears some
resemblance to that of the conventional TBA equations of the XXZ chain
\cite{TS}.  Presumably, this common structure has its origin in the
fusion rules and root of unity properties of the underlying
$U_{q}(su_{2})$ algebra.

Having in hand an exact solution of a model with so many free boundary
parameters, one can hope to be able to analyze a plethora of
interesting boundary behavior. 

Finally, we note that it should be possible to generalize the approach
presented here to open integrable anisotropic spin chains constructed
from $R$ and $K$ matrices (both trigonometric and elliptic)
corresponding to higher-dimensional representations and/or higher-rank
algebras.

We hope to be able to address some of these issues in future
publications.

\section*{Acknowledgments}

This work was supported in part by the National Science Foundation
under Grant PHY-0244261.

\begin{appendix}

\section{Appendix}\label{sec:expressions}

Here we collect explicit expressions for various quantities appearing in the
text.

The functions $\delta(u)$ and $f(u)$ are given in terms of the
boundary parameters  $\alpha_{\mp} \,, \beta_{\mp} \,, \theta_{\mp}$
by \cite{MN1}
\be
\delta(u) = \delta_{0}(u) \delta_{1}(u) \,,  \label{delta01}
\ee
where
\be
\delta_{0}(u) &=& \left( \sinh u \sinh(u + 2\eta) \right)^{2N} {\sinh 2u
\sinh (2u + 4\eta)\over \sinh(2u+\eta) \sinh(2u+3\eta)}\,, \label{delta0} \\
\delta_{1}(u) &=&  2^{4} \sinh(u + \eta + \alpha_{-}) \sinh(u + \eta - \alpha_{-})
\cosh(u + \eta + \beta_{-}) \cosh(u + \eta - \beta_{-})  \non \\
& \times & \sinh(u + \eta + \alpha_{+}) \sinh(u + \eta - \alpha_{+})
\cosh(u + \eta + \beta_{+}) \cosh(u + \eta - \beta_{+}) \,,
\label{delta1}
\ee
and therefore, 
\be
\delta_{j}(u+ i\pi) =\delta_{j}(u) \,, \qquad \delta_{j}(-u -2\eta) 
=\delta_{j}(u) \,, \qquad j = 0 \,, 1 \,.
\ee 
Moreover,
\be
f(u) = f_{0}(u) f_{1}(u) \,. \label{f01}
\ee 
For even values of $p$,
\be
f_{0}(u) &=& (-1)^{N+1} 2^{-2 p N} \sinh^{2N} \left( (p+1)u \right)
\,, \label{f0} \\
f_{1}(u) &=& (-1)^{N+1} 2^{3-2 p} \Big( \non \\
& & \hspace{-0.2in}
\sinh \left( (p+1) \alpha_{-} \right)\cosh \left( (p+1) \beta_{-} \right)
\sinh \left( (p+1) \alpha_{+} \right)\cosh \left( (p+1) \beta_{+} \right)
\cosh^{2} \left( (p+1)u \right) \non \\
&-&
\cosh \left( (p+1) \alpha_{-} \right)\sinh \left( (p+1) \beta_{-} \right)
\cosh \left( (p+1) \alpha_{+} \right)\sinh \left( (p+1) \beta_{+} \right)
\sinh^{2} \left( (p+1)u \right) \non \\
&-&
(-1)^{N} \cosh \left( (p+1)(\theta_{-}-\theta_{+}) \right)
\sinh^{2} \left( (p+1)u \right) \cosh^{2} \left( (p+1)u \right) 
\Big) \,.
\label{f1even}
\ee
For odd values of $p$,
\be
f_{0}(u) &=& (-1)^{N+1} 2^{-2 p N} \sinh^{2N} \left( (p+1)u \right)
\tanh^{2} \left( (p+1)u \right)
\,, \label{f0odd} \\
f_{1}(u) &=& -2^{3-2 p} \Big( \non \\
& & \hspace{-0.2in}
\cosh \left( (p+1) \alpha_{-} \right)\cosh \left( (p+1) \beta_{-} \right)
\cosh \left( (p+1) \alpha_{+} \right)\cosh \left( (p+1) \beta_{+} \right)
\sinh^{2} \left( (p+1)u \right) \non \\
&-&
\sinh \left( (p+1) \alpha_{-} \right)\sinh \left( (p+1) \beta_{-} \right)
\sinh \left( (p+1) \alpha_{+} \right)\sinh \left( (p+1) \beta_{+} \right)
\cosh^{2} \left( (p+1)u \right) \non \\
&+&
(-1)^{N} \cosh \left( (p+1)(\theta_{-}-\theta_{+}) \right)
\sinh^{2} \left( (p+1)u \right) \cosh^{2} \left( (p+1)u \right) 
\Big) \,. \label{f1odd}
\ee 
For both even and odd values of $p$, these functions have the properties
\be
f_{j}(u + \eta) = f_{j}(u) \,, \qquad f_{j}(-u)=f_{j}(u) \,, \qquad j = 0 \,, 1 \,.
\label{fident}
\ee

The coefficients $\mu_{k}$ appearing in the function $Y(u)$ (\ref{Y}) 
are given as follows. For $p$ even, 
\be
\mu_{0} &=& 2^{-4p} \Bigg\{ -1 - 
\cosh^{2}((p+1)(\theta_{-}-\theta_{+})) \non \\ 
&-& \cosh(2(p+1)\alpha_{-}) \cosh(2(p+1)\alpha_{+}) 
+ \cosh(2(p+1)\alpha_{-}) \cosh(2(p+1)\beta_{-}) \non \\ 
&+& \cosh(2(p+1)\alpha_{+}) \cosh(2(p+1)\beta_{-}) 
+ \cosh(2(p+1)\alpha_{-}) \cosh(2(p+1)\beta_{+}) \non \\ 
&+& \cosh(2(p+1)\alpha_{+}) \cosh(2(p+1)\beta_{+}) 
- \cosh(2(p+1)\beta_{-})  \cosh(2(p+1)\beta_{+}) \non \\ 
&+& \Big[ \cosh((p+1)(\alpha_{-}+\alpha_{+})) 
\cosh((p+1)(\beta_{-}-\beta_{+})) \non \\
&-& \cosh((p+1)(\alpha_{-}-\alpha_{+})) \cosh((p+1)(\beta_{-}+\beta_{+})) \Big]^{2}\non \\ 
&+& 2 (-1)^{N} \cosh((p+1)(\theta_{-}-\theta_{+})) \Big[
\cosh((p+1)(\alpha_{-}-\alpha_{+})) \cosh((p+1)(\beta_{-}-\beta_{+})) \non \\ 
&-& \cosh((p+1)(\alpha_{-}+\alpha_{+})) \cosh((p+1)(\beta_{-}+\beta_{+})) 
\Big] \Bigg\}
\,, \non \\ 
\mu_{1} &=& 2^{1-4p} \Bigg\{ 
\cosh((p+1)(\alpha_{-}-\alpha_{+})) 
  \Big[ \cosh((p+1)(\alpha_{-}+\alpha_{+})) \non \\
  &+& 
(-1)^{N}\cosh((p+1)(\beta_{-}+\beta_{+}))\cosh((p+1)(\theta_{-}-\theta_{+})) 
\Big] \non \\
&-& \cosh((p+1)(\beta_{-}-\beta_{+}))
\Big[ \cosh((p+1)(\beta_{-}+\beta_{+}))  \non \\
 &+& 
(-1)^{N}\cosh((p+1)(\alpha_{-}+\alpha_{+}))\cosh((p+1)(\theta_{-}-\theta_{+})) \Big]
\Bigg\} \,, \non \\ 
\mu_{2} &=& 2^{-4p} \sinh^{2}((p+1)(\theta_{-}-\theta_{+})) \,. 
\label{mueven}
\ee 
For $p$ odd,
\be
\mu_{0} &=& 2^{-4p} \Bigg\{ -1 - 
\cosh^{2}((p+1)(\theta_{-}-\theta_{+})) \non \\ 
&-& \cosh(2(p+1)\alpha_{-}) \cosh(2(p+1)\alpha_{+}) 
- \cosh(2(p+1)\alpha_{-}) \cosh(2(p+1)\beta_{-}) \non \\ 
&-& \cosh(2(p+1)\alpha_{+}) \cosh(2(p+1)\beta_{-}) 
- \cosh(2(p+1)\alpha_{-}) \cosh(2(p+1)\beta_{+}) \non \\ 
&-& \cosh(2(p+1)\alpha_{+}) \cosh(2(p+1)\beta_{+}) 
- \cosh(2(p+1)\beta_{-})  \cosh(2(p+1)\beta_{+}) \non \\ 
&+& \Big[ \cosh((p+1)(\alpha_{-}+\alpha_{+})) 
\cosh((p+1)(\beta_{-}-\beta_{+})) \non \\
&+& \cosh((p+1)(\alpha_{-}-\alpha_{+})) \cosh((p+1)(\beta_{-}+\beta_{+})) \Big]^{2}\non \\ 
&-& 2 (-1)^{N} \cosh((p+1)(\theta_{-}-\theta_{+})) \Big[
\cosh((p+1)(\alpha_{-}-\alpha_{+})) \cosh((p+1)(\beta_{-}-\beta_{+})) \non \\ 
&+& \cosh((p+1)(\alpha_{-}+\alpha_{+})) \cosh((p+1)(\beta_{-}+\beta_{+})) 
\Big] \Bigg\}
\,, \non \\ 
\mu_{1} &=&  2^{1-4p} \Bigg\{ 
\cosh((p+1)(\alpha_{-}+\alpha_{+})) 
  \Big[ \cosh((p+1)(\alpha_{-}-\alpha_{+})) \non \\
  &+& 
(-1)^{N}\cosh((p+1)(\beta_{-}-\beta_{+}))\cosh((p+1)(\theta_{-}-\theta_{+})) 
\Big] \non \\
&+& \cosh((p+1)(\beta_{-}+\beta_{+}))
\Big[ \cosh((p+1)(\beta_{-}-\beta_{+}))  \non \\
 &+& 
(-1)^{N}\cosh((p+1)(\alpha_{-}-\alpha_{+}))\cosh((p+1)(\theta_{-}-\theta_{+})) \Big]
\Bigg\} \,, \non \\ 
\mu_{2} &=&  2^{-4p} \sinh^{2}((p+1)(\theta_{-}-\theta_{+})) \,. 
\label{muodd}
\ee 

\end{appendix}

\end{document}